\documentclass[aps,amsmath,amssymb,reprint,twocolumn,pra,superscriptaddress,notitlepage,showpacs,tightenlines]{revtex4-1}
\usepackage{dcolumn}
\usepackage{amsmath}
\usepackage{amssymb}
\usepackage{bm}
\usepackage{graphicx}
\usepackage{physics}
\usepackage{pifont}
\usepackage{fancyhdr} 
\usepackage{color}
\usepackage{hyperref}
\usepackage{textcomp}

\hypersetup{hidelinks}
\hypersetup{
colorlinks=true,
linkcolor=blue,
citecolor=blue,
anchorcolor=blue,
urlcolor=blue,
}
\usepackage[toc,page]{appendix}

\newcommand{\Bracket}[1]{\left\lbrace{#1}\right\rbrace}

\begin{document}
\title{Enantiodetection of chiral molecules via two-dimensional spectroscopy}

\author{Mao-Rui Cai}
\affiliation{Beijing Computational Science Research Center, Beijing 100193, China}

\author{Chong Ye}
\affiliation{Beijing Key Laboratory of Nanophotonics and Ultrafine Optoelectronic Systems School of Physics, Beijing Institute of Technology, Beijing 100081, China}

\author{Hui Dong}
\email{hdong@gscaep.ac.cn}
\affiliation{Graduate School of China Academy of Engineering Physics, No. 10 Xibeiwang East Road, Haidian District, Beijing, 100193, China}

\author{Yong Li}
\email{liyong@csrc.ac.cn}
\affiliation{Beijing Computational Science Research Center, Beijing 100193, China}
\affiliation{Synergetic Innovation Center for Quantum Effects and Applications, Hunan Normal University, Changsha 410081, China}

\date{\today}
\begin{abstract}
  Enantiodetection of chiral molecules is important to pharmaceutical drug production, chemical reaction control, and biological function designs. Traditional optical methods of enantiodetection rely on the weak magnetic-dipole or electric-quadrupole interactions, and in turn suffer from the weak signal and low sensitivity. We propose a new optical enantiodetection method to determine the enantiomeric excess via two-dimensional (2D) spectroscopy of the chiral mixture driven by three electromagnetic fields. The quantities of left- and right- handed chiral molecules are reflected by the intensities of different peaks on the 2D spectrum, separated by the chirality-dependent frequency shifts resulting from the relative strong electric-dipole interactions between the chiral molecules and the driving fields. Thus, the enantiomeric excess can be determined via the intensity ratio of the peaks for the two enantiomers.
\end{abstract}
\maketitle

  \textit{Introduction}.\textemdash Chiral molecules contain two species (often dubbed as enantiomers), e.g. left- and right- handed chiral molecules, that are mirror images of each other but can not be superposed on through rotations, translations, or any other combinations of these two changes~\cite{mezey2012new, Struc_Dyna}. Enantiomers share almost the same physical properties, yet have disparate chemical and biological properties~\cite{mezey2012new}, which lead to constant interest on investigating enantioseparation~\cite{PhysRevLett.110.213004, doi:10.1063/1.5045052, PhysRevA.99.012513}, enantioconversion~\cite{PhysRevLett.84.1669, PhysRevA.65.015401, doi:10.1063/1.1405116,PhysRevLett.90.033001, Frishman2004, Ye_2021} as well as enantiodetection~\cite{PhysRevLett.97.173002, doi:10.1021/j100251a006, He:11, Begzjav:19}. The function of a drug with chiral mixture depends critically on the enantiomeric excess, characterizing the amount difference of the two enantiomers. Accurately determining the enantiomeric excess, i.e. enantiodetection of chiral molecules, is thus of the critical importance, yet remains a challenging task. In the traditional enantiodetection methods including circular dichroism~\cite{doi:10.1021/j100251a006}, vibrational circular dichroism~\cite{doi:10.1021/j100251a006, He:11}, and Raman optical activity~\cite{Begzjav:19}, the signal is typically weak since they depend on the weak magnetic-dipole or electric-quadrupole interactions.

  In contrast to these traditional magnetic-dipole (or electric-quadrupole) dependent methods~\cite{doi:10.1021/j100251a006, He:11, Begzjav:19}, some innovative approaches that rely only on the strong electric-dipole coupling~\cite{PhysRevA.84.053849, DoyleNature, PhysRevLett.111.023008, doi:10.1021/acs.jpclett.5b02443, doi:10.1021/jz502312t, AC-Stark, Chen-JCP, doi:10.1021/acs.jpclett.1c02196, PhysRevLett.118.123002} have been proposed and have flourished in current studies of chiral molecules. These approaches adopted a cyclic three-level structure~\cite{doi:10.1063/1.4738753, PhysRevLett.87.183002} which is constructed with three electromagnetic driving fields (nearly-) resonantly coupled to the three electric-dipole transitions of chiral molecules. In the cyclic three-level system, the overall phases of the three Rabi frequencies for the two enantiomers differ by $\pi$~\cite{PhysRevLett.87.183002, doi:10.1063/1.4738753}, resulting in chirality-dependent effective Hamiltonian and thus chirality-dependent process. Such a cyclic three-level structure offers a feasible and universal mechanism for enantiodetection~\cite{PhysRevA.84.053849, PhysRevLett.111.023008, DoyleNature, doi:10.1021/acs.jpclett.5b02443, doi:10.1021/jz502312t, AC-Stark, Chen-JCP, doi:10.1021/acs.jpclett.1c02196} (as well as enantio-specific state transfer~\cite{PhysRevLett.87.183002, PhysRevA.77.015403, Jia_2010, PhysRevLett.118.123002, https://doi.org/10.1002/anie.201704901, PhysRevLett.122.173202, doi:10.1063/1.5097406} and enantioseparation~\cite{PhysRevLett.99.130403, doi:10.1063/1.3429884, PhysRevA.104.013113}). However, many electric-dipole-transition--based enantiodetection methods typically require enantiopure samples as reference~\cite{DoyleNature, PhysRevLett.111.023008, doi:10.1021/acs.jpclett.5b02443, doi:10.1021/jz502312t}, limiting their applications since acquiring a standard enantiopure sample is also challenging~\cite{PhysRevLett.110.213004, doi:10.1063/1.5045052, PhysRevA.99.012513, PhysRevLett.84.1669, PhysRevA.65.015401, doi:10.1063/1.1405116,PhysRevLett.90.033001, Frishman2004, Ye_2021, PhysRevLett.99.130403, PhysRevA.104.013113, doi:10.1063/1.3429884} for many molecules.
  
  In this letter, we propose a new method on determining enantiomeric excess of chiral mixture via the 2D spectroscopy~\cite{mukamel1995principles, 2Dcho, shim2009, Fleming-CP2011, MIDDLETON201012} in a four-level structure of chiral molecules consisting of an upper cyclic three-level subsystem and an auxiliary level~\cite{PhysRevLett.118.123002, Chen-JCP, doi:10.1021/acs.jpclett.1c02196}. The upper cyclic three-level subsystem is formed with applying three electromagnetic driving fields, and is chirality-dependent due to the $\pi$ difference for the overall phases of the two enantiomers. The electric-dipole transition between the auxiliary level (e.g. ground state) and one of the upper three levels (e.g. the first-excited state) is coupled with three delayed laser pulses, producing a chirality-dependent signal field for the probe of the 2D spectroscopy via the process of four-wave mixing. In our scheme, the signal is detected in time domain as a function of three time coordinates ($\tau, T,$ and $t$) and Fourier transformed with respect to $\tau$ and $t$~\cite{MIDDLETON201012, Fleming-CP2011}. The signals from the two enantiomers are mapped onto the separate peaks on the spectrum with the frequency separation depending on the applied driving fields in the cyclic three-level subsystem. The relation between the intensity ratio of the peaks of interest and enantiomeric excess allows us to determine the enantiomeric excess of chiral mixture.

\begin{figure}[htbp]
    \centering
    \includegraphics[scale=0.48]{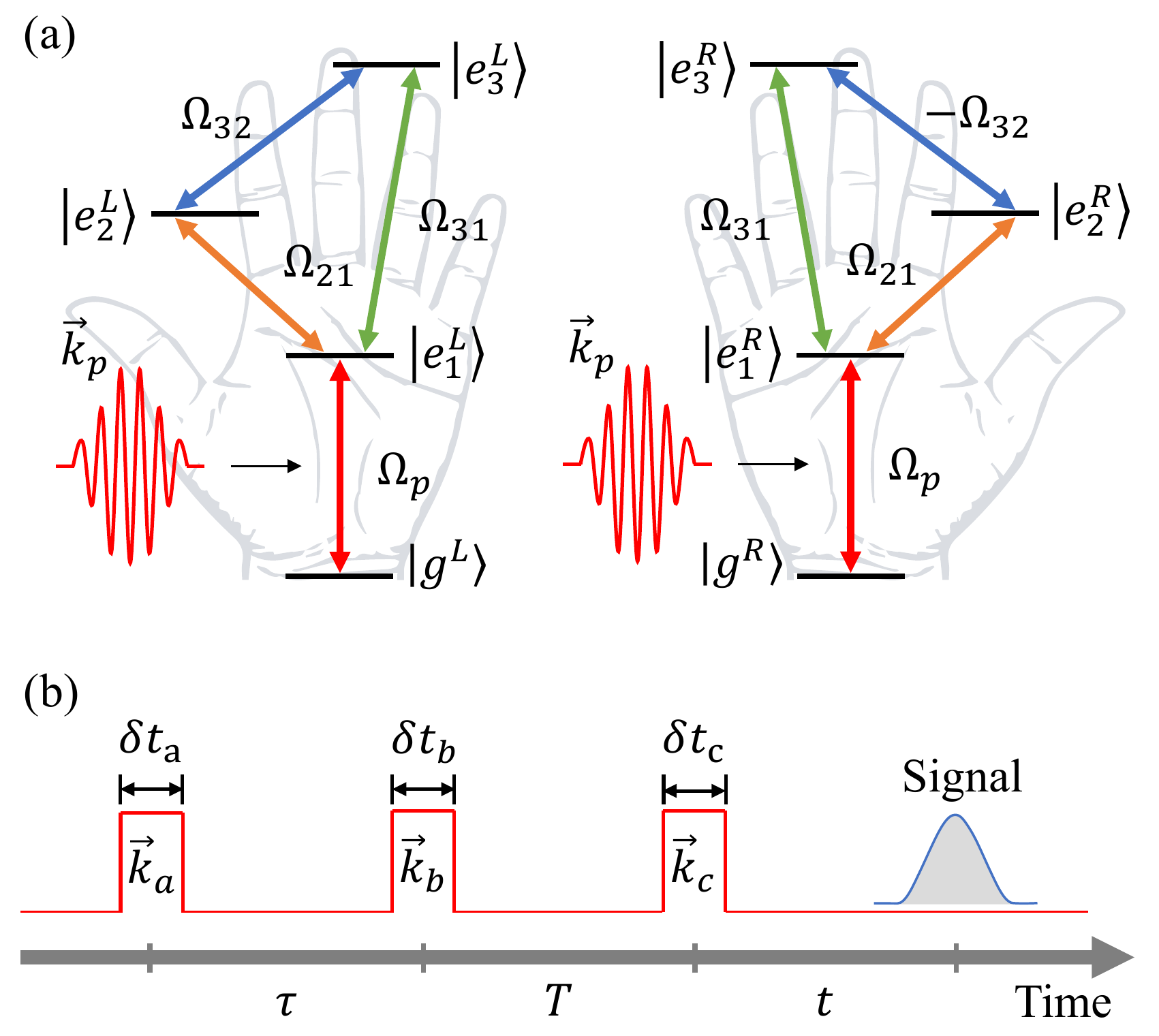}
    \caption{(Color online) The model and the pulse sequence. (a) In the cyclic three-level subsystem, three electric-dipole transitions are resonantly coupled to three electromagnetic driving fields with Rabi frequencies $\Omega_{21}$, $\Omega_{31}$, and $\pm \Omega_{32}$. The incident probe pulses with wave vectors $\vec{k}_p\:(p = a, b, c)$ are applied to induce the transition between $\ket{g^{\alpha}}$ and $\ket{e_1^{\alpha}}$ with Rabi frequencies $\Omega_p$. (b) The squares denote the incident probe pulses under the square pulse approximation, while the curve denotes the signal.}
    \label{setup}
  \end{figure}

  \textit{Model}.\textemdash The current method utilizes the generic four-level structure universally existing in chiral molecules~\cite{Chen-JCP, PhysRevLett.118.123002, doi:10.1021/acs.jpclett.1c02196}. The structure is illustrated in Fig.~\ref{setup}(a), composed of a ground state $\ket{g^{\alpha}}$ and an upper cyclic three-level subsystem with three excited states $\ket{e_1^{\alpha}}, \ket{e_2^{\alpha}},\,\text{and}\,\ket{e_3^{\alpha}}$. Here, the index $\alpha\:(= L\:\text{or}\:R)$ is used to denote the left- ($L$) or right- handed ($R$) chiral molecules. Three electromagnetic driving fields (e.g. microwave fields), shown in Fig.~\ref{setup}(a) as blue, orange, and green arrows, are constantly applied to a sample of chiral mixture to induce the chirality-dependent difference in the evolution of the two enantiomers for later probe. The frequencies of these fields are designed to be resonant with the corresponding electric-dipole transitions~\cite{AC-Stark,Chen-JCP}. In the interaction picture with respect to $H_0^{\alpha} = \sum_j \omega_j \ket{e_j^{\alpha}}\bra{e_j^{\alpha}}\:(\hbar = 1, j = 1, 2, 3)$, the Hamiltonian is given as
  \begin{equation}
    V_I^{\alpha} = \Omega_{21}^{\alpha} \ket{e_2^{\alpha}}\bra{e_1^{\alpha}} + \Omega_{31}^{\alpha} \ket{e_3^{\alpha}}\bra{e_1^{\alpha}} + \Omega_{32}^{\alpha} \ket{e_3^{\alpha}}\bra{e_2^{\alpha}} + \text{H.c.},
    \label{VI}
  \end{equation}
  where $\omega_j$ are the energies of states $\ket{e_j^{\alpha}}$ with the ground state energy $\omega_0 = 0$ (the lower index 0 corresponds to ground state hereafter), and $\Omega_{jl}^{\alpha}$ are the Rabi frequencies corresponding to the transition $\ket{e_j^{\alpha}} \leftrightarrow \ket{e_l^{\alpha}}$. The rotation-wave approximation has already been applied above. The difference between left- and right- handed chiral molecules is depicted by the Rabi frequencies~\cite{PhysRevLett.87.183002, doi:10.1063/1.4738753,AC-Stark} as $\Omega_{21}^{L}=\Omega_{21}^{R}=\Omega_{21},\Omega_{31}^{L}=\Omega_{31}^{R}=\Omega_{31}$, and $\Omega_{32}^{L}=-\Omega_{32}^{R}=\Omega_{32}$.

  The chirality dependence will be reflected by the eigenvector $\ket{d_j^{\alpha}} =\sum_{i} n_{ij}^{\alpha} \ket{e_{i}^{\alpha}}$ ($i, j = 1, 2, 3$) of the cyclic three-level subsystem~\cite{AC-Stark} under the Hamiltonian $V_{I}^{\alpha}$, which satisfies $V_{I}^{\alpha}\left|d_{j}^{\alpha}\right\rangle =E_{j}^{\alpha}\left|d_{j}^{\alpha}\right\rangle $ with $E_{j}^{\alpha}$ being the corresponding eigenvalues. Here, $n_{ij}^{\alpha}$ are the transformation matrix elements. The detailed derivations are presented in the Supplementary Material. Our current method is designed to determine the signal from different enantiomers by distinguishing the frequencies $\text{\ensuremath{E_{j}^{\alpha}}}$.

  Three laser pulses (e.g. infrared pulses) in the BoxCARS geometry~\cite{mukamel1995principles, 2Dcho, shim2009, Fleming-CP2011, MIDDLETON201012} with intervals $\tau$ and $T$ are applied to probe the response of the sample. $\vec{k}_p$ $(p = a, b, c)$ are the wave vectors and $\delta t_p$ are the pulse durations for these pulses. The central frequencies $\nu$ of the three pulses are set to be the same and resonant with the transition $\ket{g^{\alpha}} \leftrightarrow \ket{e_1^{\alpha}}$, i.e. $\nu = \omega_1$. The Hamiltonian describing the current system during a pulse interaction is written in the interaction picture as
  \begin{equation}
      V_p^{\alpha}(s) = \Omega_p(s) e^{i \vec{k}_p \cdot \vec{r}} \ket{e_1^{\alpha}}\bra{g^{\alpha}} + \text{H.c.},
    \label{Vp}
  \end{equation}
  where $\vec{r}$ is the spatial location of the molecule, $\Omega_p(s)$ is the Rabi frequency corresponding to the transition $\ket{g^{\alpha}} \leftrightarrow \ket{e_1^{\alpha}}$, and $s$ is the time variable. Within the short durations of the probe pulses, we have neglected the interaction Hamiltonian $V_I^{\alpha}$ in Eq.~(\ref{Vp}) noticing that the incident pulses are strong and short enough that the interaction with the three driving fields becomes negligible, i.e. $\int_{0}^{\delta t_{p}}\Omega_{jl}ds\ll\int_{0}^{\delta t_{p}}\Omega_{p}(s)ds$.

  As shown by the pulse sequence in Fig.~\ref{setup}(b), the sample of chiral mixture interacts with the first pulse $\vec{k}_a$ and then interacts with the second pulse $\vec{k}_b$ after an evolution time $\tau$. Later, after a second evolution time $T$, the sample interacts with the third pulse $\vec{k}_c$, while the signal field emitted by the sample will be measured after a third evolution time $t$. After the whole process, the final wave function of the system is given (in the Schr\"{o}dinger picture) as 
    \begin{equation}
      \ket{\psi^{\alpha}(\tau, T, t)}
      = U_S^{\alpha}(t) U_{c,S}^{\alpha} U_S^{\alpha}(T) U_{b,S}^{\alpha} U_S^{\alpha}(\tau) U_{a, S}^{\alpha} \ket{\psi_0^{\alpha}}
      \label{finalstate}
    \end{equation}
  with $\ket{\psi_0^{\alpha}}$ being the initial state, which is taken as the ground state $\ket{g^{\alpha}}$ under the typical low-temperature condition~\cite{DoyleNature}. Here, $U_{p, S}^{\alpha} \equiv \exp[-i H_0^{\alpha} \delta t_{p}]U_{p}^{\alpha}(\delta t_{p})$ and $U_{S}^{\alpha}(s)\equiv\exp[-i H_{0}^{\alpha} s]\exp[-iV_{I}^{\alpha}s]$ are, respectively, the evolution operators with and without the probe pulses. The evolution operator $U_{p}^{\alpha}(\delta t_{p})\equiv \mathcal{T} \exp[-i\int_{0}^{\delta t_{p}}V_{p}^{\alpha}(s)ds]$ (with $\mathcal{T}$ being the time-ordering operator) is simplified under the square pulse approximation~\cite{Chen_2019} as $U_{p}^{\alpha}(\delta t_{p}) = \exp[-i V_p^{\alpha} \delta t_p]$. Here $V_p^{\alpha} = \Omega_p \exp[i \vec{k}_p \cdot \vec{r}] \ket{e_1^{\alpha}}\bra{g^{\alpha}} + \text{H.c.}$ is the interaction Hamiltonian during square pulse $\vec{k}_p$ and $\Omega_p$ is the corresponding (time-independent) Rabi frequency.

  The emission of the two chiral molecule is induced by the polarization $\bm{P}^{\alpha}(\tau, T, t) = \text{Tr}\left[\rho^{\alpha}(\tau, T, t) \bm{\mu}^{\alpha}\right]$, where $\rho^{\alpha}(\tau,T,t) \equiv \ket{\psi^{\alpha}(\tau,T,t)} \bra{\psi^{\alpha}(\tau,T,t)}$ is the density matrix and $\bm{\mu}^{\alpha}$ is the electric-dipole operator. We particularly sort out the signal with the frequencies near $\omega_1$, which is proportional to $\bm{P}_{10}^{\alpha}(\tau,T,t) \equiv \rho_{10}^{\alpha}(\tau,T,t) \bm{\mu}_{01} + \text{c.c.}$ (with $\bm{\mu}_{01}^{\alpha} = \bm{\mu}_{01}$ being the electric-dipole moment corresponding to the transition $\ket{e_{1}^{\alpha}} \rightarrow \ket{g^{\alpha}}$). Moreover, by the phase matching, we select the (\textit{rephasing}) signal emitted along the direction $\vec{k}_{s}=-\vec{k}_{a}+\vec{k}_{b}+\vec{k}_{c}$ as $\bm{P}_{10}^{\alpha, \text{RP}}(\tau,T,t)\exp[i\vec{k}_s\cdot\vec{r}]$, where
  \begin{equation}
    \begin{aligned}
    \bm{P}_{10}^{\alpha, \text{RP}}(\tau,T,t)=\, & \bm{\mu}_{01} \sum_{l,l' = 1}^3 \left|n_{1l}^{\alpha}\right|^2 \left|n_{1l'}^{\alpha}\right|^2\times\\
    \Big{[} & Ae^{i(\omega_{1}+E_{l}^{\alpha})(\tau+T)}e^{-i(\omega_{1}+E_{l'}^{\alpha})(T+t)}\\
    + & Be^{i(\omega_{1}+E_{l}^{\alpha})\tau}e^{-i(\omega_{1}+E_{l'}^{\alpha})t}\Big{]}.
    \end{aligned}
    \label{signal}
  \end{equation}
  Here $A=\mathcal{N}_{a}^{2}\mathcal{N}_{b}^{2}\mathcal{N}_{c}\beta_{a}^{*}\beta_{b}\beta_{c}^{*}$, and $B=\mathcal{N}_{a}^{2}\mathcal{N}_{b}\mathcal{N}_{c}^{2}\beta_{a}^{*}\beta_{b}^{*}\beta_{c}$, $\beta_p \simeq -i \Omega_p \delta t_p$ under the condition $\Omega_p \delta t_p \ll 1$ and $\mathcal{N}_p = (1 + \left|\beta_p\right|^2)^{-1/2}$ are the renormalized constants. And the index RP means the \textit{rephasing} signal.

  \begin{figure}
    \centering 
    \includegraphics[scale=0.35]{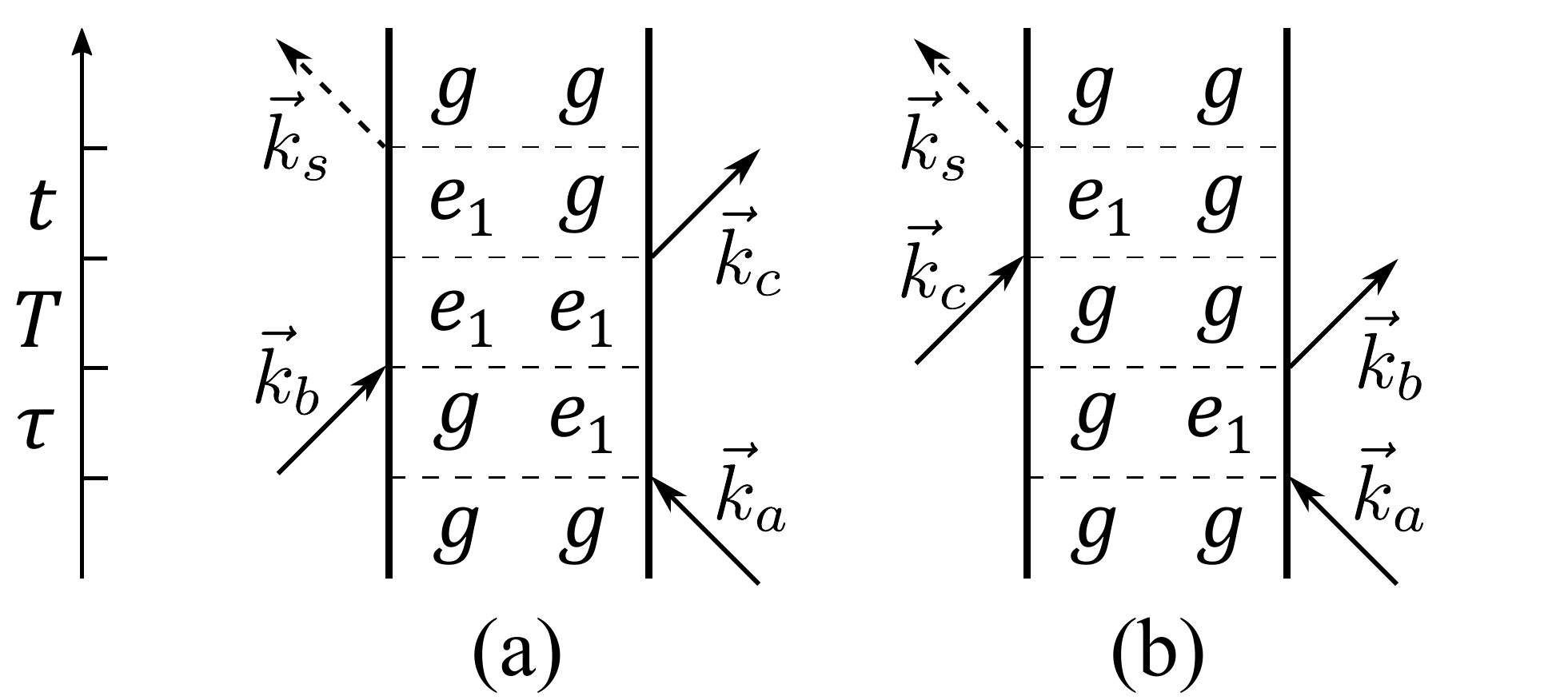}
    \caption{Double-side Feynman diagrams for (a) stimulated emission process and (b) ground state bleach process with the evolution times being denoted on the left side. The solid and dash arrows in the two diagrams represent the incident probe pulses and the signals, respectively. In (a) the stimulated emission process, the population for the first-excited state is generated after the first two pulses, while in (b) the ground state bleach process, the ground-state population is generated.}
    \label{FeynmanDiag}
  \end{figure}

  The two terms in Eq.~(\ref{signal}) correspond to two different processes: the stimulated emission (SE) and the ground state bleach (GSB), illustrated by the double-side Feynman diagrams in Fig.~\ref{FeynmanDiag}. The two processes differ from each other at the duration time $T$, i.e. in the SE process the excited-state population is generated while in the GSB process the ground-state population is generated. Taking into account the relaxation rate $\Gamma$ and the pure dephasing rate $\gamma$ of the excited states (assumed to be the same for three excited states) in these processes, the \textit{rephasing} signal in Eq.~(\ref{signal}) is thus corrected as
  \begin{equation}
    \begin{aligned}
    \bm{P}_{10}^{\alpha, \text{RP}} & (\tau,T,t)=\bm{\mu}_{01}\sum_{l,l' = 1}^3 \left|n_{1l}^{\alpha}\right|^2 \left|n_{1l'}^{\alpha}\right|^2\times\\
    \Big{[} & Ae^{i(\omega_{1}+E_{l}^{\alpha})(\tau+T)}e^{-i(\omega_{1}+E_{l'}^{\alpha})(T+t)}e^{-\Gamma'(\tau+t)}e^{-\Gamma T}\\
    + & Be^{i(\omega_{1}+E_{l}^{\alpha})\tau}e^{-i(\omega_{1}+E_{l'}^{\alpha})t}e^{-\Gamma'(\tau+t)}\Big{]},
    \end{aligned}
    \label{finalsignal}
  \end{equation}
  where $\Gamma'\equiv\frac{\Gamma}{2}+\gamma$ indicates the decay of $\rho_{10}$.

  For a chiral mixture with $N_L$ left-handed molecules and $N_R$ right-handed molecules, the average \textit{rephasing} signal is
  \begin{equation}
    \begin{aligned}
      \bm{P}_{10}^{m, \text{RP}}&(\tau, T, t) = \frac{N_L \bm{P}_{10}^{L, \text{RP}} (\tau, T, t)
      + N_R \bm{P}_{10}^{R, \text{RP}} (\tau, T, t) }{N_L + N_R} ,
    \end{aligned}
    \label{mixSig}
  \end{equation}
  where the upper index $m$ means chiral mixture.

  \textit{2D spectrum}.\textemdash The measured signal in time domain is analyzed~\cite{Fleming-CP2011} with the 2D Fourier transform with respect to $\tau$ and $t$ as
  \begin{equation}
    \tilde{\bm{P}}_{10}^{m, \text{RP}}(\omega_{\tau},T,\omega_{t}) \equiv \bm{\mathcal{F}} \left[\bm{P}_{10}^{m, \text{RP}}(\tau,T,t)\right],
    \label{spectrum}
  \end{equation}
  which is the frequency-domain spectrum. By substituting Eqs.~(\ref{finalsignal}) and (\ref{mixSig}) into Eq.~(\ref{spectrum}), it is clear that (for the general case of non-degenerated eigenvalues $E_j^{\alpha}$) each enantiomer will generate nine peaks at positions $(\omega_{\tau}, \omega_t) = (\omega_1 + E_l^{\alpha}, -\omega_1 - E_{l'}^{\alpha})$ in the 2D spectrum with their magnitudes dominated by $|n_{1l}^{\alpha}|^2 |n_{1l'}^{\alpha}|^2$. The chirality dependence of $E_j^{\alpha}$ and $n_{1j}^{\alpha}$ will thus be reflected by the spectrum.

  \begin{figure*}[htbp]
    \centering
    \includegraphics[scale=0.43]{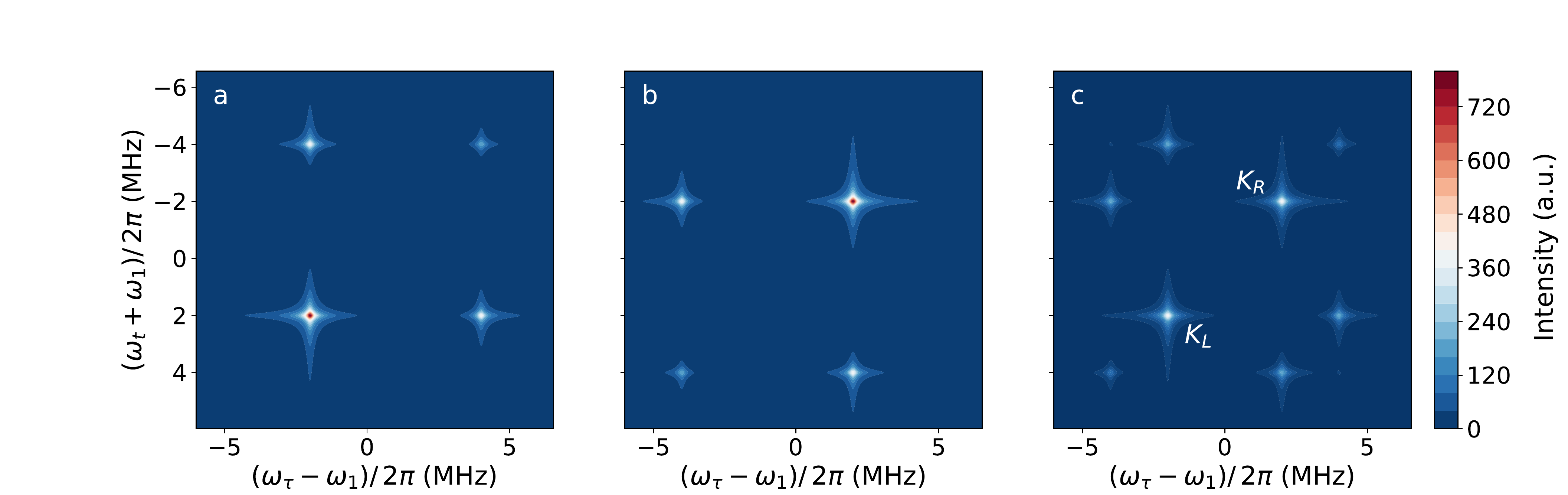}
    \caption{(Color online) 2D spectra of (a) pure left-handed chiral molecules, (b) pure right-handed chiral molecules, and (c) racemic mixture with equal left- and right- handed chiral molecules. These spectra are obtained via 2D fast Fourier transform and only the absolute values of the transform result are taken. The time-domain signals are scanned from 0 to 0.2 $\text{ms}$ with 0.02 $\mu \text{s}$ time step for both $\tau$ and $t$, while $T$ is set to be zero here.}
    \label{2d}
  \end{figure*}

  We demonstrate our method on determining enantiomeric excess by an example of 1,2-propanediol with equal Rabi frequencies of the cyclic three-level subsystem, i.e. $\Omega_{21} = \Omega_{31} = \Omega_{32} = \Omega$~\cite{PhysRevLett.90.033001}. Under this arrangement, the eigenvalues of the Hamiltonian $V_I^L$ (for left-handed chiral molecules) are $E_1^L = 2\Omega$, $E_2^L = E_3^L = -\Omega$, and the eigenvalues of the Hamiltonian $V_I^R$ (for right-handed chiral molecules) are $E_1^R = -2\Omega$, $E_2^R = E_3^R = \Omega$. The transformation matrix elements involved in Eq.~(\ref{finalsignal}) are also specified as $|n_{1j}^{\alpha}|^2 = 1/3$.

  To give a numerical simulation, we choose the working states as $\ket{g} = \ket{v_g}\ket{0_{0, 0, 0}}$, $\ket{e_1} = \ket{v_e} \ket{1_{1, 1, 1}}$, $\ket{e_2} = \ket{v_e}\ket{2_{2, 1, 2}}$, and $\ket{e_3} = \ket{v_e}\ket{2_{2, 0, 1}}$, where the chirality index is neglected. The ket vector $\ket{v_g}$ $(\ket{v_e})$ denotes the corresponding vibrational ground (first-excited) state and $\ket{J_{K_a, K_c, M}}$ denotes the rotational state~\cite{AngularM}. The transition frequencies are given as $\omega_{10} / 2\pi \simeq 4.24\,\text{THz}$, $\omega_{21} / 2\pi \simeq 29.21\,\text{GHz}$, $\omega_{31} / 2\pi \simeq 29.31\,\text{GHz}$, and $\omega_{32} / 2\pi \simeq 100.76\,\text{MHz}$~\cite{AngularM, ARENAS20179, LOVAS200982,Chen-JCP}, where $\omega_{ij} = |\omega_i - \omega_j|$. We take the Rabi frequency $\Omega / 2\pi \simeq 2\,\text{MHz}$, the relaxation rate $\Gamma / 2\pi \simeq 1\,\text{kHz}$, and the pure dephasing rate $\gamma / 2\pi\simeq 0.1\,\text{MHz}$ according to the current experiments~\cite{doi:10.1080/00268976.2012.679632}. We assume that all the incident probe pulses are the same except for their direction by taking $\Omega_p / 2\pi \simeq 50\,\text{MHz}$ and $\delta t_p \simeq 0.5\,\text{ns}$, with the bandwidth $\delta \nu \simeq 2 \pi \times 0.9\,\text{GHz} \ll \Bracket{\omega_{21}, \omega_{31}}$~\cite{DIELS20061}.
  
  The numerical results are shown in Fig.~\ref{2d} where the time-domain signals are detected every 0.02\,$\mu \text{s}$ with the scale $0.2\,\text{ms}$ for time coordinates $\tau$ and $t$. The 2D spectra are obtained via 2D fast Fourier transform by taking the absolute values of the transform result. To obtain a maximum response, we have taken $T = 0$ in Fig.~\ref{2d}, since the signal should decay with relaxation rate $\Gamma$ during the evolution time $T$, as indicated by Eq.~(\ref{finalsignal}). However, if the time coordinate $T$ is also scanned, one is able to discriminate $\Gamma$ out of $\Gamma'$, which could be a further application of our method.

  Due to the degeneracy for the eigenvalues $E_j^{\alpha}$ in the above case, nine peaks reduce to four (two diagonal and two off-diagonal peaks) for each enantiomer in Figs.~\ref{2d}(a) and \ref{2d}(b). For the chiral mixture, there are eight peaks (four peaks for left-handed and four peaks for right-handed chiral molecules) in total. And with the help of the off-diagonal peaks, one can directly determine which four correspond to the same enantiomer while a standard enantiopure sample is not needed. 

\begin{figure}
    \centering
    \includegraphics[scale=0.37]{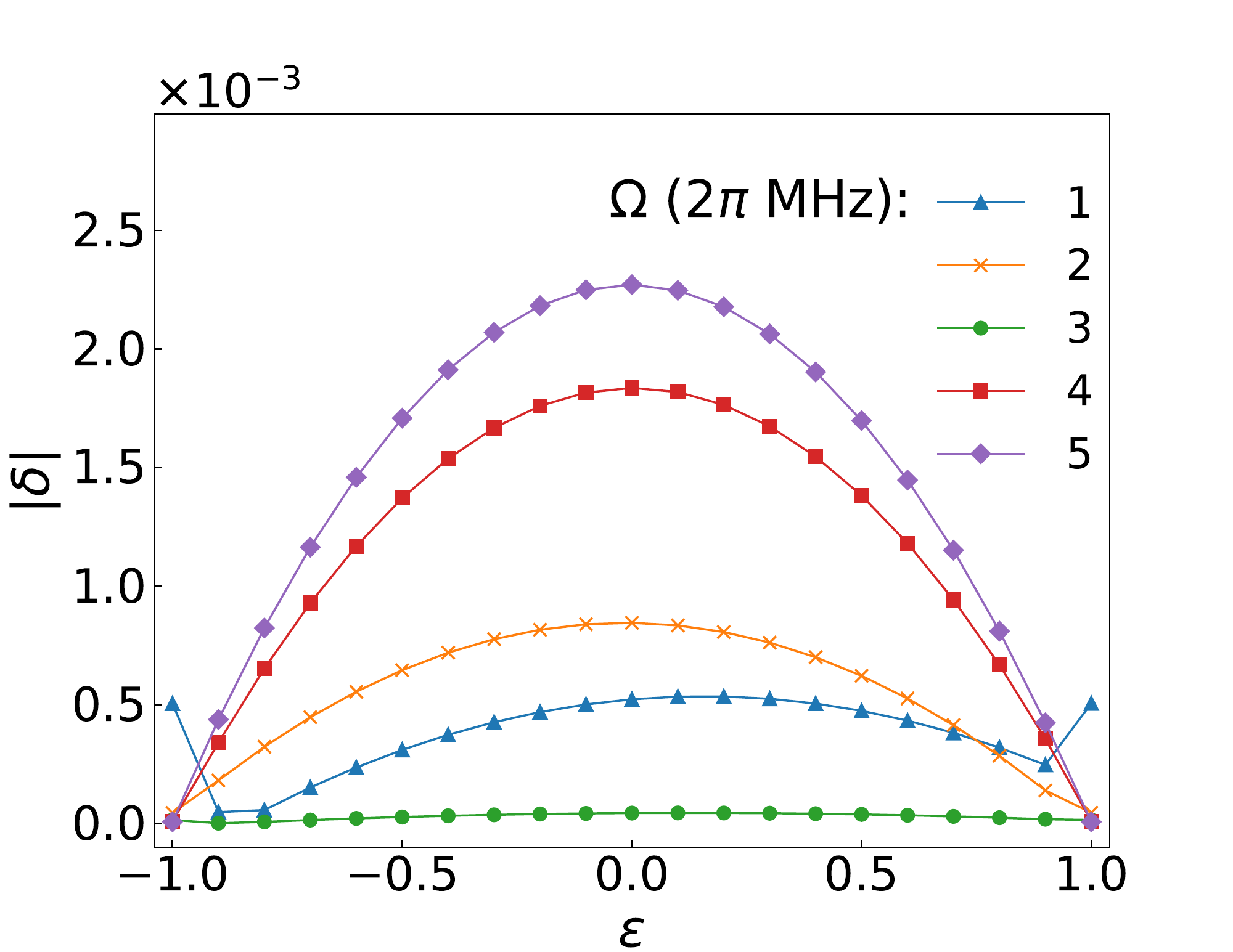}
    \caption{(Color online) The absolute errors of our estimation on the enantiomeric excess for different Rabi frequencies. The other parameters are adopted from Fig.~\ref{2d}. Moreover, the time domain signals are also scanned with the same path in Fig.~\ref{2d}.}
    \label{ee}
  \end{figure}

  To determine the enantiomeric excess, we denote $K_L$ ($K_R$) the amplitude of the highest peak for left-handed (right-handed) chiral molecules  [e.g. see Fig.~\ref{2d}(c) in the case of racemic mixture with $N_L = N_R$], 
  \begin{equation}
    \begin{aligned}
      &K_{L/R} = (N_L + N_R) \times \left|\bm{\tilde{P}}_{10}^{m, \text{RP}} (\omega_{\mp}, T, -\omega_{\mp})\right|, \\
      &= N_L \left|\bm{\tilde{P}}_{10}^{L, \text{RP}} (\omega_{\mp}, T, -\omega_{\mp})\right| 
      + N_R \left|\bm{\tilde{P}}_{10}^{R, \text{RP}} (\omega_{\mp}, T, -\omega_{\mp})\right|
    \end{aligned}
  \end{equation}
  (the indices $``-"$ and $``+"$ on the right-hand side correspond to indices $L$ and $R$ on the left-hand side, respectively), where $\omega_{\mp} = \omega_1 \mp \Omega$. In the strong coupling condition $\Bracket{\Gamma, \gamma} \ll \Omega$, the peaks are well separated, and $K_L$ ($K_R$) is thus approximately linearly proportional to $N_L$ ($N_R$),
  \begin{equation}
    \begin{aligned}
      K_L \simeq N_L \left|\bm{\tilde{P}}_{10}^{L, \text{RP}} (\omega_{-}, T, -\omega_{-})\right|, \\
      K_R \simeq N_R \left|\bm{\tilde{P}}_{10}^{R, \text{RP}} (\omega_{+}, T, -\omega_{+})\right|. 
    \end{aligned}
  \end{equation}
  Noticing the equal coefficients $|n_{1j}^{\alpha}|^2 = 1/3$ in Eq.~(\ref{finalsignal}) in the case of degenerate eigenvalues, the signal intensities for left- and right- handed chiral molecules are the same, i.e. $|\bm{\tilde{P}}_{10}^{L, \text{RP}} (\omega_{-}, T, -\omega_{-})| = |\bm{\tilde{P}}_{10}^{R, \text{RP}} (\omega_{+}, T, -\omega_{+})|$. Then, our estimation of the enantiomeric excess is given as
  \begin{equation}
    \varepsilon_e = \frac{K_L - K_R} {K_L + K_R},
    \label{est_ee}
  \end{equation}
  and the error of our estimation is $\delta = \varepsilon_e - \varepsilon$, with $\varepsilon = (N_L - N_R) / (N_L + N_R)$ being the real enantiomeric excess. Fig.~\ref{ee} shows the effectiveness of our estimation by numerically giving the absolute errors for various Rabi frequencies $\Omega$. It reveals that in most region the absolute errors of our estimation are smaller than $2 \times 10^{-3}$. 

  We note that our estimation is not limited to the condition where $|n_{1j}^{\alpha}|^2$ are equal for both enantiomers under the degenerate-eigenvalue case with the arrangement $\Omega_{21} = \Omega_{31} = \Omega_{32} = \Omega$. For the general case when $\Omega_{21}$, $\Omega_{31}$, and $\Omega_{32}$ are not equal, one can still denote $K_L$ ($K_R$) the amplitude of the highest peak for left-handed (right-handed) chiral molecules, and $K_{\alpha} \simeq N_{\alpha} |\bm{\tilde{P}}_{10}^{\alpha, \text{RP}} (\omega_{\alpha}, T, -\omega_{\alpha})|$. Here, $\omega_L = \omega_1 + E_{\xi}^L$ and $\omega_R = \omega_1 + E_{\zeta}^R$ ($\xi, \zeta \in \{1, 2, 3\}$), which decide the locations of the two chosen peaks, are experiment-specified. According to Eq.~(\ref{finalsignal}), the signal intensities for left- and right- handed chiral molecules have the relation $|\bm{\tilde{P}}_{10}^{L, \text{RP}} (\omega_L, T, -\omega_L)| / |n_{1\xi}^{L}|^4 = |\bm{\tilde{P}}_{10}^{R, \text{RP}} (\omega_R, T, -\omega_R)| / |n_{1\zeta}^{R}|^4$. Thus, for the general case, the estimation of the enantiomeric excess in Eq.~(\ref{est_ee}) is corrected as $\varepsilon_e = (K_L |n_{1\zeta}^R|^4 - K_R |n_{1\xi}^{L}|^4) / (K_L |n_{1\zeta}^{R}|^4 + K_R |n_{1\xi}^{L}|^4)$.

  \textit{Conclusion}.\textemdash We have proposed a new method on enantiomeric excess determination via the 2D spectroscopy based on the generic four-level structure of chiral molecules. The three driving fields cause the different frequency shifts of the upper three levels for the two enantiomers. By introducing three additional laser pulses, the signal generated by the four-wave-mixing process is probed via the 2D spectrum, where the peaks reflect the chirality dependence of the shifts. With the existence of the off-diagonal peaks, the peaks that correspond to the same enantiomer are naturally identified. Thus, our proposal does not require the standard enantiopure sample. In the strong coupling condition $\Bracket{\Gamma, \gamma} \ll \Omega$, the peaks for the two enantiomers are well separated and their amplitudes are approximately linear with the quantities of the corresponding chiral molecules. Hence, the enantiomeric excess of the chiral mixture can be estimated by comparing the peak intensities with low error.

  The advantages of the current method lie in two aspects. Firstly, the tunable driving fields allow the creation of significant difference between enantiomers, comparing with the traditional enantiodetection methods with the weak magnetic-dipole or electric-quadrupole interactions. Secondly, the application of 2D spectroscopy allows the separate contributions of the enantiomers within one single spectrum, without the need for standard enantiopure sample, benefited from the inherit resolution of 2D spectroscopy.
  
  We acknowledge discussions with Y.-Y. Chen. This work was supported by the National Natural Science Foundation of China (under Grants No.~12074030, No.~12088101, No.~11875049, No.~U1930402, and No.~U1930403).
  
\bibliography{Chiral2DSpectrumCMR}
\end{document}